\documentclass[3p,times]{elsarticle}

\usepackage{ecrc}

\usepackage{epstopdf}

\volume{00}

\firstpage{1}

\journalname{Nuclear Physics A}

\runauth{Miklos Gyulassy et al.}

\jid{nupha}

\jnltitlelogo{Nuclear Physics A}

\usepackage{graphicx}
\usepackage{amsmath,amssymb}

\newcommand{\br}{bremsstrahlung }

\begin{document}

\begin{frontmatter}



\title{Initial-State Bremsstrahlung versus Final-State Hydrodynamic
Sources of Azimuthal Harmonics in $p+A$ at RHIC and LHC}

\author[label1,label2]{M. Gyulassy}
\author[label1]{P. Levai}
\author[label3]{I. Vitev}
\author[label1]{T.~S.~Bir\'o}

\address[label1]{MTA WIGNER Research Centre for Physics, RMI, Budapest, Hungary }
\address[label2]{Department of Physics, Columbia University, New York, 
10027, USA}
\address[label3]{Theoretical Division, Los Alamos National Laboratory, Los
Alamos, NM 87545, USA} 

\begin{abstract}
Recent $p_T<2$~GeV azimuthal correlation data 
from the Beam Energy Scan (BES) and D+Au runs at RHIC/BNL and, especially, 
the surprising 
similarity of azimuthal $v_n\{2m\}(p_T)$ ``transeverse flow'' harmonics
in $p+Pb$ and $Pb+Pb$ at LHC have challenged the uniqueness of local equilibrium
``perfect fluid'' interpretations of those data. 
We report results derived in~\cite{GLVB14} on azimuthal harmonics
associated with initial-state non-abelian ``wave interference'' 
effects predicted by 
perturbative QCD gluon bremsstrahlung and sourced by Color Scintillation Arrays (CSA) 
of color antennas. CSA are naturally identified with 
multiple projectile and target {\em beam jets} produced in inelastic 
$p+A$ reactions.  We find a remarkable similarity between
 azimuthal harmonics     
sourced by initial state CSA and those predicted with final state 
perfect fluid models of high energy  $p+A$ reactions. The question of which mechanism dominates in $p+A$ and $A+A$ remains open at this time.   
\end{abstract}


\end{frontmatter}


\section{Introduction}
An unexpected discovery at RHIC/BNL in $D+Au$ reactions at $\sqrt{s}=200$~AGeV~\cite{Adare:2013piz}   and 
at LHC/CERN  in $\sqrt{s}=5.02$~ATeV $p+Pb$  reactions~\cite{CMS:2012qk}
is the large magnitude of mid-rapidity 
azimuthal anisotropy moments, $v_n{2}(k_T, \eta=0)$, that are remarkably 
similar to those observed previously in non-central $Au+Au$
~\cite{Adams:2005dq}
and in $Pb+Pb$
~\cite{Aamodt:2011by}
reactions. 
See especially the preliminary ATLAS results in Fig. 24 of ref.~\cite{ATLASvn}
 and the QM14 ATLAS talk by J. Jia in these proceedings~\cite{ATLASpAv1}.
In addition, the Beam Energy Scan (BES) at RHIC~\cite{Adamczyk:2013gw}
revealed a near $\sqrt{s}$ independence from 8~AGeV to 2.76~ATeV of
$v_n(p_T;s)$ in $A+A$ at fixed centrality that was also unexpected. 

In high energy $A+A$, the systematics of $v_n\{2\ell\}(k_T,\eta=0)$ data 
have been interpreted as
possible evidence for the near ``perfect fluidity'' of the
strongly-coupled Quark Gluon Plasmas (sQGP) produced in such
reactions~\cite{Romatschke:2007mq, 
Heinz:2013th}.  However, the recent observation of similar $v_n$ arising
from much smaller transverse size $p(D)+A$ systems and also the near beam energy
independence of the moments observed in the Beam Energy Scan (BES)~\cite{Adamczyk:2013gw}
from 7.7~AGeV to 2.76~ATeV in $A+A$ have posed a problem for the
perfect fluid interpretation because near inviscid hydrodynamics is
not expected to apply in space-time regions where the local
temperature field falls below the confinement temperature, $T(x,t) <T_c\sim
160 $~MeV. In that Hadron Resonance Gas (HRG) ``corona'' region, the
viscosity to entropy ratio is predicted to grow rapidly with
decreasing temperature~\cite{Danielewicz:1984ww} and the corona volume
fraction must increase and the volume of the
perfect fluid ``core'' with $T>T_c$ must decrease 
when either  the projectile atomic
number $A$ decreases or the center-of-mass (CM) energy
$\sqrt{s}$ decreases .

While hydrodynamic equations have been shown to be  
{\em sufficient}  to describe $p(D)+A$ data
with particular assumptions about initial and freeze-out boundary
 conditions~\cite{Bozek:2011if}, its {\em necessity} as a 
unique interpretation of the data is not guaranteed. 
This point was underlined recently using a specific initial-state 
saturation model~\cite{Dusling:2013oia}  that was shown to be able to fit  $p(D)+A$ 
correlation $v_{2n}$ even moments data  without final-state interactions.  
That saturation model has also been used in~\cite{Gale:2012rq} to specify
initial conditions for perfect fluid hydrodynamics in  $A+A$.  
However, in $p+A$  the transverse spatial structure
of initial conditions is not as well-controlled
 because the gluon saturation scale scale, $Q_s$, is small and its fluctutations
over the transverse plane  on sub-nucleon scales are more model dependent than in $A+A$.

The near independence of $v_n$ moments on beam energy observed in the BES~\cite{Adamczyk:2013gw}
at RHIC from 7.7 AGeV to 2.76~ATeV pose another serious challenges to
the uniqueness of the perfect fluid  interpretations of the data
because previous hybrid fluid-HRG modeling~\cite{Teaney:2000cw}  predicted a natural systematic reduction of the moments due to the increasing HRG corona fraction with decreasing beam energy.
 The  HRG corona fraction is expected to dilute flow signatures from the perfect fluid QGP core flow at lower energies. 
The BES~\cite{Adamczyk:2013gw} data also a pose a challenge  to color glass condensate (CGC)
gluon saturation models~\cite{Kharzeev:2004bw}  used to  specify initial conditions for hydrodynamic 
flow  predictions in  $A+A$. This is because $Q_s^2$ is  predicted to decrease with
$\log (s)$, and, thus, gluon saturation-dominated ``central rapidity region'' gluon fusion dynamics must switch 
over into valence quark-diquark dominated ``fragmentation region'' inelastic 
dynamics involving fragmentation of multiple quark-diquark beam jets.

In our GLVB paper~\cite{GLVB14} we explored the  possibility that initial-state gluon bremsstrahlung,
sourced by  Color Scintillating Arrays (CSA)  of colored beam jet antennae, could partially 
account for the above puzzling systematics of azimuthal harmonics. Non-abelian \br
is intrinsically azimuthally anisotropic. The pQCD-based GLVB model extends 
the first order in opacity $\chi=1$  Gunion-Bertsch~\cite{Gunion:1981qs} 
(GB) perturbative QCD \br 
to all orders  in opacity, $e^{-\chi}\sum_{n=1}^\infty\chi^n/n! \cdots$~\cite{Gyulassy:2000er}, 
Vitev-Gunion-Bertsch (VGB)~\cite{Vitev:2007ve}  multiple interaction pQCD \br 
for applications to $B+A$ nuclear collisions. 
We show that VGB \br naturally leads on an 
event-by-event basis to a hierarchy of 
non trivial azimuthal asymmetry moments remarkably similar to that observed in 
$p+A$ and peripheral $A+A$ at fixed $dN/d\eta$~\cite{Chatrchyan:2012wg}

\section{Results}
\begin{figure}[!tbh]
\centerline{\includegraphics[width=3.in]{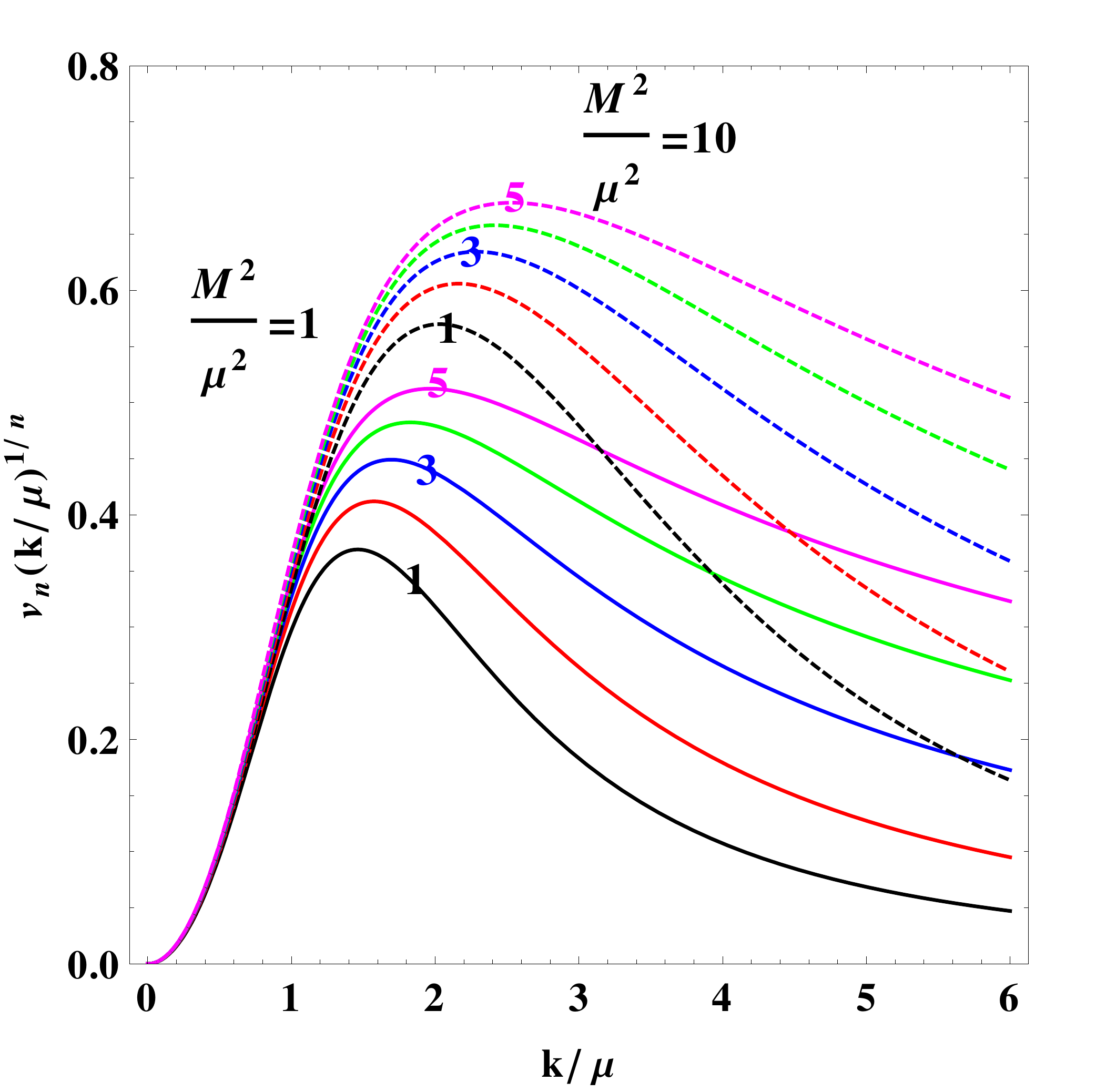}
\hspace{0.1in} \includegraphics[width=3.5in]{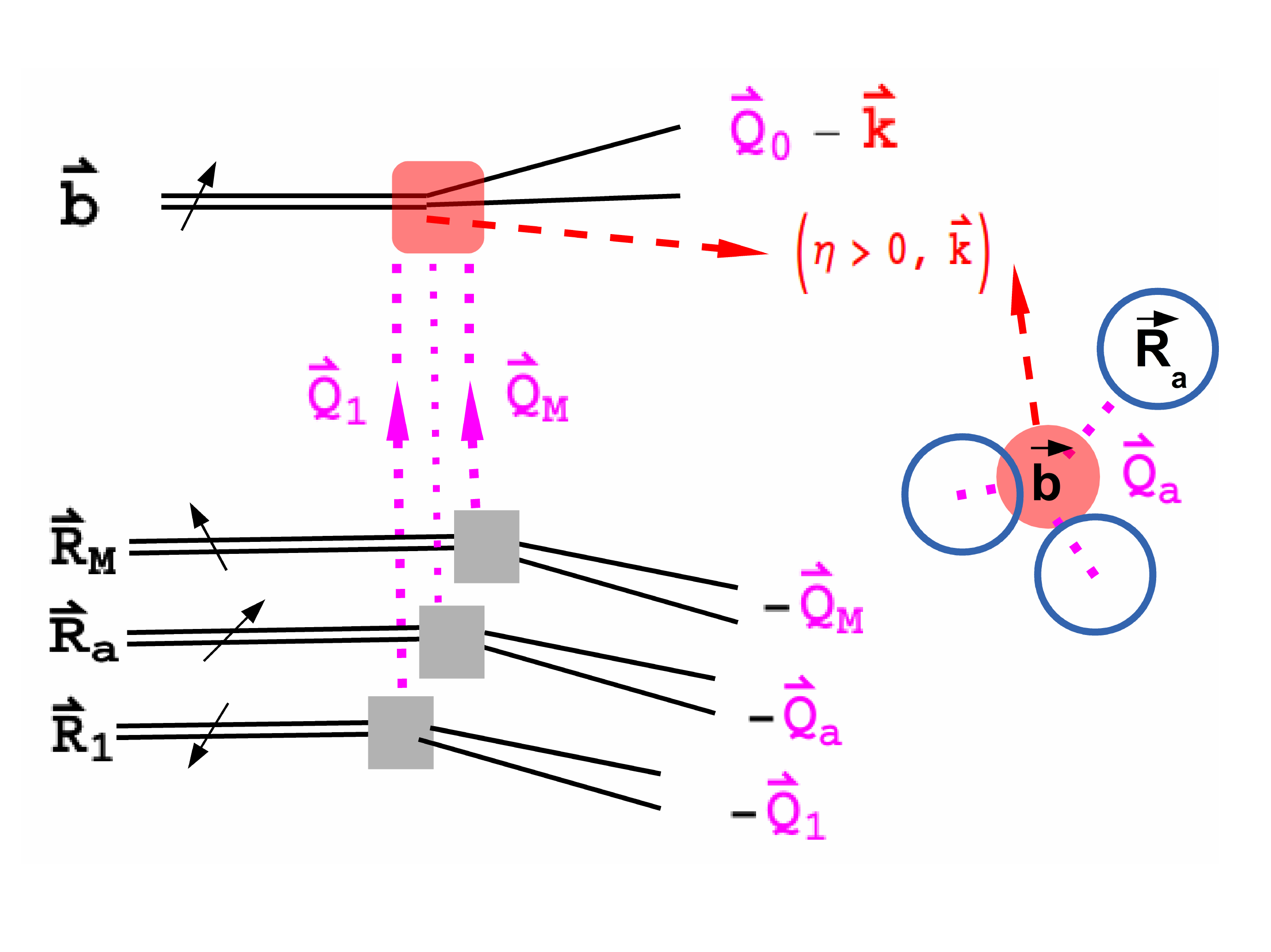}
}
\caption{(Color online) (a) [left] Approximate $1/n$ power scaling
of $q$ averaged $ \langle v_n^{GB}(k,q,0)\rangle^{1/n}$ for a fixed reaction
plane $\psi=0$  from a single GB color antenna for all even and odd moments
is seen in the kinematic region  
 $k^2 <  \langle q^2\rangle\equiv M\mu^2$.  This scaling  breaks down for for
$k >  \sqrt{M}\mu$ 
because in the $\mu=0$ limit \br stricly vanish for $k>q$. (b) [right]
Schematic diagram corresponding 
to coherent \br from the projectile beam jet at impact parameter ${\bf b}$. 
The azimuthal distribution 
is enhanced for transverse momenta ${\bf k}$ near the total accumulated
momentum transfer to the projectile 
${\bf Q}_0\equiv \sum_{a} {\bf Q}_a$ from the $a=1,\cdots, M$
clusters of recoiling target beam jets. In addition, the \br is enhanced near
${\bf k}=-{\bf Q}_a$ from each incoherent recoiling target cluster 
separated by $\Delta R_{ab} \sim 1/k$
in the transverse plane.  
}
\label{fig-4new}
\end{figure}

The non-abelian bremsstrahlung Gunion-Bertsch (GB) 
formula~\cite{Gunion:1981qs,Vitev:2007ve} for the soft gluon radiation single inclusive 
distribution for a triggered beam jet recoil momentum ${\bf q}=(q,\psi)$ is  
\begin{eqnarray}
 \frac{dN_g^1}{d\eta d^2 {\bf k}d^2{\bf q}}  &= &  \frac{C_R \alpha_s}{\pi^2} 
\frac{\mu^2}{\pi(q^2+\mu^2)^2}
\frac{{\bf q}^2}{{\bf k}^2 ({\bf k}-{\bf q})^2}   
\;\; .  \label{GB1} \end{eqnarray}
Here, the 
parton scattering elastic cross section is assumed to be
$d\sigma_{0}/d^2{\bf q}=\sigma_{0} \mu^2/\pi(q^2+\mu^2)^2$. The produced gluon has rapidity $\eta$ and 
transverse momentum ${\bf k}=(k,\phi)$. Note especially that 
the azimuthally asymmetric angular dependence has the simple form, 
$dN_g= F_{kq}/(A_{kq} - \cos(\phi-\psi))$   of 
the radiated $\phi$ relative to the reaction plane $\psi$ angles arising from
basic non-abelian interference effects.  Note also the uniform rapidity-even, $\eta \approx \log (x E/k)$, 
distribution of non-abelian bremsstrahlung. 
In $p+A$ multiple target beam jets generally transform that uniform $\eta$ dependence into 
a trapezoidal one, as discussed in~\cite{GLVB14}. 
The GB azimuthal harmonics can then be analytically evaluated from
\begin{eqnarray}
\hspace{-0.2in} 
v_n^{GB}(k,q,\psi)
&=&  \int\frac{ d\phi}{2\pi} \cos(n \phi)\; \frac{(A_{kq}^2-1)^{1/2}}{ A_{kq} - \cos(\phi-\psi)}
 \;\;=\;\; \cos[n \psi] \;(\; v_1^{GB}(k,q)\;)^n \; ,\label{gbsc}\\
v_1^{GB}(k,q) &=& (A_{kq}-\sqrt{A^2_{kq}-1})
\; \; , \label{vn1}
\end{eqnarray}
where (see \cite{GLVB14})  $A_{kq}= (k^2+q^2+\mu^2)/(2 k \, q) \ge 1$ implies that all harmonics
are peaked near $k\sim q$, vanish at $k=0$, and slowly decrease toward zero for $k\gg q$. In addition, the analytic single color antenna GB gluon harmonics
obey an approximate power law scaling with respect to the harmonic $n$ number:
\begin{eqnarray}
[v_n^{GB}(k,q,0)]^{1/n} &=& \;[v_m^{GB}(k,q,0)]^{1/m}
\; \; ,\label{vnscaling}\end{eqnarray}
that is similar to the scaling observed by ALICE, CMS and 
ATLAS~\cite{Aamodt:2011by
} at LHC
and similar to perfect fluid harmonic scaling  for the higher $n\ge 3$ moments 
dominated by purely geometric fluctuations. We note also that unlike the low order CGC azimuthal harmonics,
the GB \br harmonics are non-vanishing and scale for all odd as well as even
moments $n$. We illustrate in Fig.~\ref{fig-4new}a the main features of azimuthal harmonics from 
a single beam jet bremsstrahlung. 

In a given $p+A$ event a projectile
nucleon penetrates through a target nucleus $A$  at impact parameter
${\bf b}$, producing one projectile beam jet with  rapidity $Y_P>0$.
There are $N\sim A^{1/3} <A$  target beam jets with  $Y_T<0$  produced
with  transverse  coordinates ${\bf R}_i$ distributed within a distance $\sim 1/\mu$ 
from the impact parameter. The $N$ target dipoles naturally cluster
near the projectile  impact parameter ${\bf{b}}$ as illustrated in 
Fig.~(1b). 
In a specific event,  there are  however  only  $1\le M\le N$ 
overlapping clusters that can radiate coherently toward 
the negative rapidity $\eta<0$ hemisphere (see \cite{GLVB14}). 
 Incoherence of target clusters
\br is controlled by the transverse resolution scale 
with $|{\bf R}_i-{\bf R}_j|> 1/k$. 

Let $ {\bf Q}_a=\sum_{i\in I_a} {\bf q}_{i}$ denote the cumulative momentum transfer to the 
projectile from target cluster $a$. 
The total single inclusive  \br  distribution into a particular 
mode $({\bf k}_1)$ has the form
\begin{equation}
 dN^{M,N}= dN^{N}_{P}(\eta,{\bf k}_1 ; {\bf Q}_P)
+dN^{M,N}_{T}(\eta,{\bf k}_1;\{{\bf Q}_{a}\})
= \sum_{a=0}^{M}  
\frac{ B_{k_1 Q_a}}{ ({\bf k}_1+{\bf Q}_a)^2+\mu_a^2 } 
\;\; , 
\label{GBPTM} 
\end{equation}
where we define 
${\bf Q}_0 = -\sum_a {\bf Q}_a$ to 
include the projectile beam jet contribution into the summation over target clusters. Note that for a semi exclusive  event with all $M$ target recoil
momenta $Q_a$ and their azimuthal orientation $\psi_a$ determined, the \br
radiation is peaked near the $M+1$ cumulative momenta. However, averaging over all reaction planes forces all single particle $v_n{1}$ to vanish on the average.
Only 2 or higher particle correlation can reveal the  intrinsic azimuthal anisotropy correlations above. Fortunately, in this CSA \br model 
all $2\ell$ relative 
azimuthal harmonics can also be evaluated analytically (see ~\cite{GLVB14}).

In the ``mean recoil'' approximation $Q\approx \bar{Q}$,
we find that a single GB antenna satisfies the generalized power scaling
law in case that subsets of the $2\ell$ gluons
have identical momenta. Suppose  there are $1\le L\le 2\ell$
distinct momenta $K_r$  with $r=1,\cdots,L$ such 
$m_r$ of the $2\ell$ gluons have  momenta equal 
to a particular value $K_r$ such that $\sum_{r=1}^{L} m_r= 2\ell$.
In this case 
$
v_n^{M=1}\{2\ell\}(k_1,\cdots,k_{2\ell};\bar{Q} )
\approx   \prod_{r=1}^L (v_n^{GB}(K_r,\bar{Q}))^{m_r}
=  \prod_{r=1}^L 
(v_1^{GB}(K_r,\bar{Q}))^{n m_r}$.
The  approximate  factorization and power scaling of azimuthal harmonics 
from CSA coherent state non-abelian \br  is similar
to ``perfect fluid hydrodynamic collective flow'' factorization and scaling,
but in this case no assumption about local equilibration or minimal viscosity 
is necessary.  
  
\section{Conclusions}

In this talk we summarized from Ref.~\cite{GLVB14}
some of the remarkable azimuthal correlation properties
of beam jet non-abelian \br even at the lowest order of perturbative QCD level
using the VGB generalization~\cite{Vitev:2007ve} 
of GB~\cite{Gunion:1981qs} \br to all orders in opacity in $p+A$.
Of course, higher order and especially high gluon occupation number effects~\cite{Kharzeev:2004bw}
could modify the intricate initial-state chromo wave interference patterns. 
However, the main lesson from this study is that in $p+A$  initial-state 
wave interference phenomena may well dominate over any final-state
dynamics but appear as if ``perfect fluid'' or ``conformal holographic''  descriptions\cite{Basar:2013hea}  were applicable on sub nucleon transverse scales.
Our analysis 
shows that long range in $\eta$ multi-gluon, $1/n$ power law scaling, 
azimuthal multipole 
cumulant signatures are not unique to final-state perfect fluid
and dual AdS shock wave flows but can arise also 
naturally from  perturbative QCD features of initial state
 \br from Color Scintillating Arrays of multiple beam jets.
A possible way to help discriminate between
initial-state interference harmonics and  final-state flow harmonics
may be through the study of rapidity dependence of multi-gluon azimuthal harmonics
 as discussed in~\cite{GLVB14}.  

\section{Acknowledgements}
 MG acknowledges support from the 
US-DOE DE-FG02-93ER40764,  DE-AC02-05CH1123. PL, TB, and MG
acknowledge support from Hungarian OTKA grants  K104260,  NK106119, and
NIH TET\_12\_CN-1-2012-0016. IV was supported in part by the US Department of Energy, 
Office of Science, Office of Nuclear Physics.

\end{document}